\documentclass[onecolumn,epsfig]{elsart}
\usepackage{epsfig}

\usepackage{psfig}
\usepackage{epsfig}
\usepackage{floatflt}
\usepackage{graphicx}
\usepackage{dcolumn}
\usepackage{bm}

\newcommand{\AmS}{{\protect\the\textfont2A\kern-.1667em\lower.5ex\hbox{M}\kern-.125emS}}
\def\Journal#1#2#3#4{{#1} {\bf #2}, #3 (#4)}

\def\NIMA{{ Nucl. Instrum. Methods} A}

\def\NPA{{ Nucl. Phys.} A}
\def\PLB{{ Phys. Lett.}  B}
\def\PRL{ Phys. Rev. Lett.}

\def\PRC{{ Phys. Rev.} C}
\def\PRD{{ Phys. Rev.} D}

\def\JPG{{ J. Phys.} G}

\def\be{\begin{equation}}
\def\ee{\end{equation}}
\def\bea{\begin{eqnarray}}
\def\eea{\end{eqnarray}}

\begin{document}
\begin{frontmatter}
\title{Extracting the charm cross-section from semileptonic decays into muons}

\author[bnl,ustc,lbl]{Haidong~Liu}
\author[ustc,lbl]{Yifei~Zhang}
\author[bnl,siap]{Chen~Zhong}
\author[bnl]{Zhangbu~Xu}
\address[bnl]{Brookhaven National Laboratory, Upton, New York 11973}
\address[ustc]{University of Science \& Technology of China, Anhui
230027, China} 
\address[lbl]{Lawrence Berkeley National Laboratory,
Berkeley, California 94720} 
\address[siap]{Shanghai Institute of Applied
Physics, Shanghai 201800, P.R. China}

\date{\today}
\begin{abstract}
We propose a sensitive measurement of the charm total cross-section at
RHIC through muon identification from charm semileptonic decay at low
transverse momentum ($p_T$).  This can test the binary-collision
scaling ($N_{bin}$) properties of the charm total cross-section and be
used to study whether heavy-flavor quarks, which are used as a probe,
are produced exclusively at the initial impact in hadron-hadron
collisions. The effect of the charm semileptonic decay form factor on
extracting the total charm cross section and on the shape of the
lepton spectra are also discussed in detail. We conclude that lepton
spectra from charmed hadron decays at transverse momentum
$p_T\simeq1.0$ GeV/$c$ are sensitive to the charmed hadron spectrum
shape. Therefore, the interactions of heavy quarks with the medium
created in relativistic heavy-ion collisions, especially the flow
effects, can be extracted from the lepton spectra from charmed hadron
decays at low $p_T$.
\end{abstract}
\end{frontmatter}

In relativistic heavy-ion collisions, charm quarks are believed to be
produced in the early stage via initial gluon fusion and their
production cross-section can be evaluated using perturbative
QCD~\cite{cacciari}.  Study of the $N_{bin}$ scaling properties of the
charm total cross-section in p+p, d+Au and Au+Au collisions can test
if heavy-flavor quarks, which are used as a probe, are produced
exclusively at the initial impact. The interactions of heavy quarks
with the medium provide a unique tool for probing the hot and dense
matter created in ultra-relativistic heavy-ion collisions at the early
times.  At RHIC energies, heavy quark energy loss~\cite{dokshitzer01},
charm quark coalescence~\cite{loic,pbm,rafelski,mclerran}, the effect
of $J/\psi$ production from charm quark coalescence on the
interpretation of possible $J/\psi$ suppression due to color
screening~\cite{matsui}, and charm flow~\cite{xu2,teaney,batsouli}
have been proposed as important tools in studying the properties of
matter created in heavy ion collisions. The last three effects depend
strongly on the charm total cross-section and spectrum at low $p_T$.

However, it is difficult to directly reconstruct charmed hadrons and
single electrons from charm semileptonic decay in hadron-hadron
collisions with high precision at low $p_{T}$, where the yield
accounts for a large fraction of the total cross
section~\cite{stardAucharm,CDF,phenixAuAu}. The difficulties are due
to large combinatorial backgrounds in charmed hadron decay channels,
and the overwhelming photon conversions in the detector material, and
$\pi^{0}$ Dalitz decays in electron detection. Nevertheless, the charm
total cross-sections have been measured in d+Au collisions at RHIC by
a combination of the directly reconstructed low $p_T$ $D^0\rightarrow
K\pi$ and the non-photonic electron spectra~\cite{stardAucharm}, and
by electron spectra alone at $p_T>0.8$
GeV/$c$~\cite{phenixAuAu,phenixpp}. Although the systematic and
statistical errors are large, the result indicates a much larger charm
yield than predicted by pQCD
calculations~\cite{stardAucharm,cacciari}.  Recently, the measurements
of high $p_T$ electrons from heavy-flavor semileptonic decays have
posed challenges to our understanding of partonic energy loss in the
medium~\cite{phenixAuAu,starcharmQM05,miklos,xindongQM05}.  Since most
of the measurements to date at RHIC are from indirect heavy-flavor
semileptonic decays, it is therefore important to find novel
approaches to improve the measurements and also study in detail how
to extract the maximum information about the heavy-flavor spectrum from its
lepton spectrum. This includes studying the effects of the form
factors of charmed hadron semileptonic decays on the lepton
spectra~\cite{pdgcharmff}.

In this paper, we propose a new method to extract the charm total
cross-section by measuring muons from charmed hadron semileptonic
decay at low $p_T$ (e.g. $0.16^{<}_{\sim}p_T{}^{<}_{\sim}0.26$
GeV/$c$).  Since muons in this $p_T$ range are a very uniform
sample of the whole charmed hadron spectrum, the inferred charm total
cross-section is insensitive to the detail of the charm spectrum.
Once the cross-section is determined, the electron spectrum at higher
$p_T$ can be used to sensitively infer the charmed hadron spectral
shape.  Meanwhile, we survey the form factors used in charm
semileptonic decays generated from Particle Data
Group~\cite{pdgcharmff}, in the PYTHIA event 
generator~\cite{stardAucharm,phenixAuAu,phenixpp,pythia}, by
pQCD predictions~\cite{cacciari} and from the CLEO inclusive
measurement~\cite{cleocetalk}.  We find that the lepton spectra from
these different form factors can be different by a factor of 1.5.

\begin{figure}
  \includegraphics[width=5in]{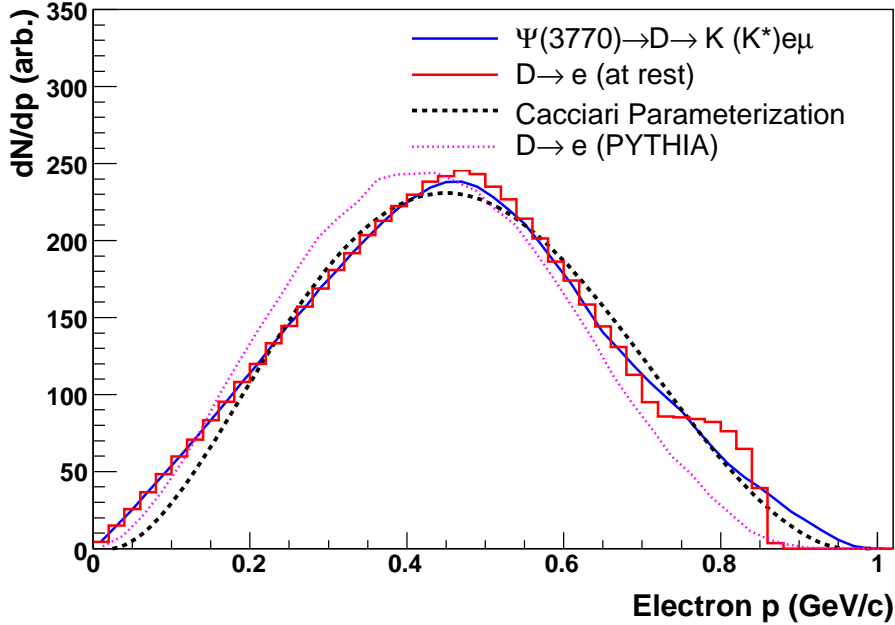}
  \caption{Electron momentum spectra from charmed meson decays at 
rest.  Histogram is the one with form factor from PDG. The 
dashed line is that of Cacciari's parameterization.  The dotted line 
is from the simplified vector-meson form factor in PYTHIA.  The solid line 
is that from the PDG which takes a form factor from
$\Psi$(3770)$\rightarrow D\rightarrow e$. 
}
  \label{fig:eatrest}
\end{figure}

We first study the charmed hadron semileptonic decay form factor and
its effect on the lepton spectrum. Fig.~\ref{fig:eatrest} shows the
electron momentum spectra from charmed meson decays at rest generated
using the Particle Data Table~\cite{pdgcharmff},
PYTHIA~\cite{stardAucharm,phenixAuAu,phenixpp,pythia}, pQCD
calculations~\cite{cacciari} and from the CLEO preliminary inclusive
measurement~\cite{cleocetalk}. The spectrum generated by the PDG is
according to the form factor of charmed meson decays to pseudoscalar
$K+l+\nu$, vector meson $K^*+l+\nu$ and non-resonance $(K\pi)+l+\nu$
where the $K^*$ mass is used for the $(K\pi)$ system.  The decay partial
widths ($\Gamma$) of the three dominant decay channels are:
\begin{enumerate}
\item $K+l+\nu$ with pseudo scalar meson in final state ($D^{\pm}$ 7.8\%)\\
$${{d\Gamma}\over{dq^{2}}} \propto
{{p_{k}^{3}}\over{(1-q^{2}/M^{*2})^2}}$$ where $q^{2}$ is the
invariant mass of the virtual $W\rightarrow\nu l$, $p_K$ is the
momentum of the kaon, and $M^{*}=0.189$ GeV/$c^2$ is a parameterization of
the effective pole mass in the decay.
\item $(K\pi)+l+\nu$ with non-resonant $K\pi$ in final state($D^{\pm}$ 4.0\%)\\
We use the $K^*$ mass (0.892 GeV/$c^2$) for the $K\pi$ invariant mass  
and the form factor is the same as in the decay to the pseudo scalar meson. 
\item $K^*+l+\nu$ with vector meson in final state($D^{\pm}$ 5.5\%)\\
$${{d\Gamma}\over{dq^{2}d\cos{\theta_{l}}}} \propto {{p_Vq^{2}}\over{M^2}} 
[(1-\cos{\theta_l})^2|H_{+}(q^{2})|^2+
{4\over3}(1+\cos{\theta_l})^2|H_{-}(q^{2})|^2+
{8\over3}\sin^2{\theta_l}|H_{0}(q^{2})|^2]$$\\

where $\theta_l$ is the decay angle between the lepton and the vector meson, 
$p_V$ is the vector meson momentum, 

$$H_{\pm}(q^{2})= (M+m)A_{1}(q^{2})\mp{{2Mp_V}\over{M+m}}V(q^{2})$$ 

and 

$$H_{0}(q^{2}) = {1\over{2mq}}[(M^2-m^2-q^{2})(M+m)A_{1}(q^{2})-
{{4M^{2}p_{V}^{2}}\over{M+m}}A_{2}(q^{2})]$$ 

where $A_{1,2},V$ take the form of $1/(1-q^{2}/M^{*2}_{A,V})$ with
$M^*_{A}=2.5$ GeV/$c^2$, $M^*_{V}=2.1$ GeV/$c^2$ and $r_V =
V(0)/A_{1}(0)=1.62\pm0.08$, $r_2 =
A_{2}(0)/A_{1}(0)=0.83\pm0.05$~\cite{pdgcharmff,polemass}.
\end{enumerate}
For different charmed hadrons, we assume that the relative branching
ratios among these three channels are the same, and their decay
electron spectra are the same. The overall charmed hadron to electron
branching ratio $\Gamma(c\rightarrow e)/\Gamma(c\rightarrow anything)$
is 10.3\%~\cite{pdgcharmff}. There is a possible $\sim5\%$ difference
between electron and muon decays due to phase space which was not
taken into account in this analysis. Electrons at high momentum are
mainly from decay channel (1) $K+l+\nu$ because the kaon is lighter
than the $K^*$ and the form factor of the decay channel to a vector
meson ($K^*+l+\nu$) favors a low momentum lepton and higher momentum
neutrino. Since PYTHIA uses a simplified vector meson decay form
factor~\cite{pythia}, it tends to produce a softer electron
spectrum. Both the parameterization by Cacciari~\cite{cacciari} and
formulae from the PDG agree with CLEO's preliminary electron
spectrum. In addition, we also find that although the charmed mesons
($D^{\pm}$ and $D^{0}$) from $\Psi$(3770) decay have a momentum of 244
MeV/c only and without correction of final state
radiation~\cite{cleocetalk}, it affects slightly its subsequent
electron spectrum.

\begin{figure}
  \includegraphics[width=5in]{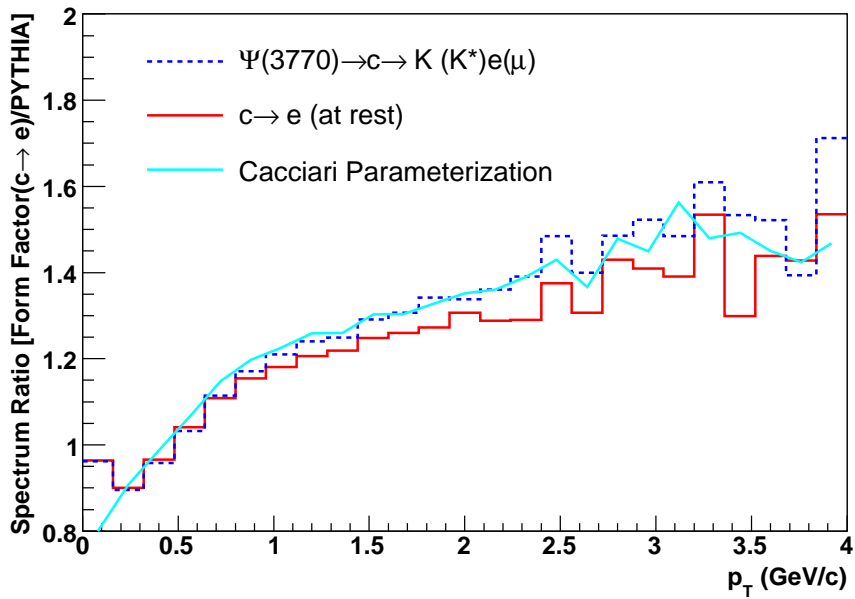}
  \caption{Charm-decay electron spectra for three different form
factors divided by the spectrum using the PYTHIA form factor. The
histogram represents the ratio using the PDG form factor. The solid
line is Cacciari's parameterization, and the dashed line is from PDG
and takes a form factor from $\Psi$(3770)$\rightarrow D\rightarrow
e$. See text for detail.}
  \label{fig:eratio2pythia}
\end{figure}

We use our electron momentum spectra and that of Cacciari to generate
 electron spectra from charmed decay at RHIC. A power-law function of
 the charmed hadron transverse momentum spectrum was obtained from
 minimum-bias Au+Au collisions~\cite{starcharmQM05}.
 Fig.~\ref{fig:eratio2pythia} shows the ratios of those electron
 spectra divided by the spectrum using PYTHIA decay form
 factors~\cite{pythia,stardAucharm}. The slightly soft form factor of
 the charm semileptonic decay in PYTHIA convoluted with a steeply
 falling charm spectrum produces an electron $p_T$ spectrum in Au+Au
 collisions at RHIC, which can be lower than the correct one by up to
 a factor of 1.5 at high $p_T$. Part of the discrepancy between
 experimental results and PYTHIA in electron
 spectra~\cite{stardAucharm,phenixAuAu} can be explained by the decay
 form factor. Taking $D^{\pm}$ and $D^{0}$ from $\Psi$(3770) decay as
 if it were at rest, results in slight change on the electron spectrum.

Since the parameters for the form factor from the PDG are extracted
from experimental data, and the form factor describes data well, we
will select this form factor with D at rest, to generate lepton
spectra for the remainder of the paper unless otherwise specified. A
power-law function is used to create charmed hadron $p_T$ spectra. The
function takes the form:
$$ {{dN}\over{2\pi dyp_Tdp_T}}={{dN}\over{dy}}
{{2(n-1)(n-2)}\over{\pi(n-3)^{2}\langle p_{T}
\rangle^{2}}}(1+{{2p_T}\over{\langle p_{T} \rangle(n-3)}})^{-n},
$$ where {\it dN/dy} is the yield and $n$ and $\langle p_{T} \rangle$
are the parameters controlling the shape of the spectrum.
\begin{figure}
  \includegraphics[width=5in]{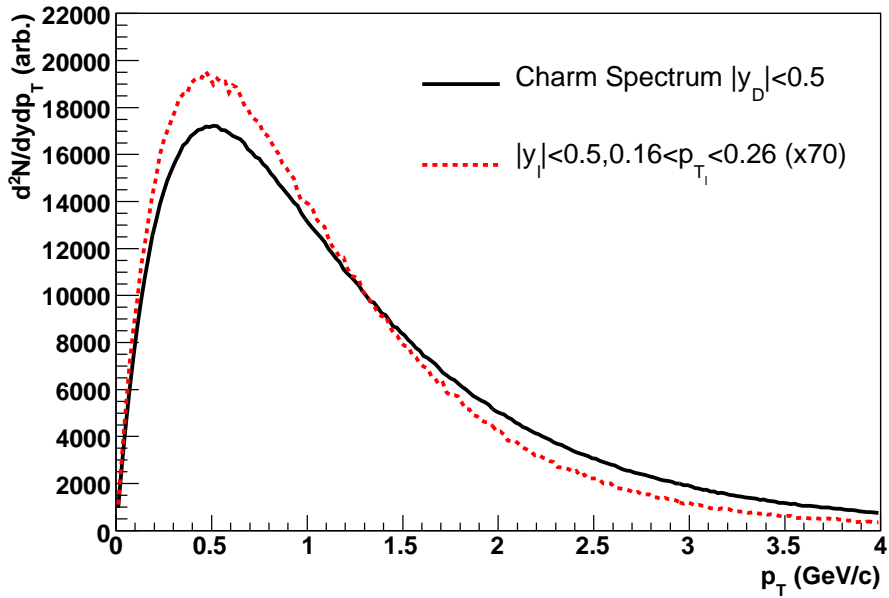}
  \caption{Charmed hadron ($D$) spectra ($dN/dydp_T$) as a function of
$p_T$ at midrapidity before (solid line) and after (dashed line) a
muon selection of $0.16<p_T<0.26$ GeV/$c$ and $|y_l|<0.5$. The later
was scaled up by a factor of 70.}
  \label{fig:D2ept}
\end{figure}
Fig.~\ref{fig:D2ept} shows the charmed hadron $p_T$ spectrum before
and after requiring its decayed muons at $0.16<p_T<0.26$ GeV/$c$. The
similarity of the spectral shape shows that the muon selection
reasonably uniformly samples the entire charmed hadron spectrum.  The
muons in this $p_T$ range sample 14\% of the charmed hadron spectrum.
The muon yield is about $1/70$ of the charmed yield due to an
additional 10.5\% decay branching ratio.  Fig.~\ref{fig:muonnmpt}
shows the dependence of the muon yield on $\langle p_{T} \rangle$ for
a fixed total charm yield. The yield is normalized to yields at $n=10$
and $\langle p_{T} \rangle=1.3$ GeV/$c$.  We also note that the muon
yield has a very weak dependence on $n$ which demonstrates that over a
wide range in $\langle p_{T} \rangle$, the muon yield is within
$\pm15\%$. This is in contrast to the large variation of the electron
yield integrated above $p_T$ of 1.0 GeV/$c$, where a factor of 8
variation is seen in Fig.~\ref{fig:muonnmpt}. This may explain the
difference between results from Au+Au, p+p and d+Au
collisions~\cite{xuISMD}, pointing to a softer charm spectrum in Au+Au
collisions. On the other hand, once {\it dN/dy} is determined by the
low $p_T$ muons, the electron yield at higher $p_T$ is very sensitive
to the $\langle p_{T} \rangle$ of the charmed hadron spectrum. If the
lepton spectrum at low $p_T$ can be measured, the electron yield at
higher $p_T$ essentially determines the charm $\langle p_{T}
\rangle$. In fact, the dependence on $\langle p_{T} \rangle$ is almost
linear. At lower beam energies at RHIC, the electron background is too
high to make a meaningful charm measurement~\cite{pidNIMA} due to the
overwhelming photon conversions in the detector material and
background from $\pi^{0}$ Dalitz decays. The alternative muon
measurement can be used to extract the charm total cross-section and
study the energy excitation function of charm
production~\cite{stardAucharm,cacciari}. In central Au+Au collisions,
$c\rightarrow\mu$ in this $p_T$ range is estimated to be about 0.1 per
event due to the high production of charm quarks at RHIC. If the
charm-decay muon yields are verified to follow binary collision
scaling in p+p, d+Au, minimum-bias and central Au+Au collisions, we
will be able to use it as an experimental measure of $N_{bin}$ for
peripheral collisions where Glauber model
calculations\cite{starwhitepaper} result in a large uncertainty on 
$N_{bin}$.

\begin{figure}[ht]
  \includegraphics[width=5in]{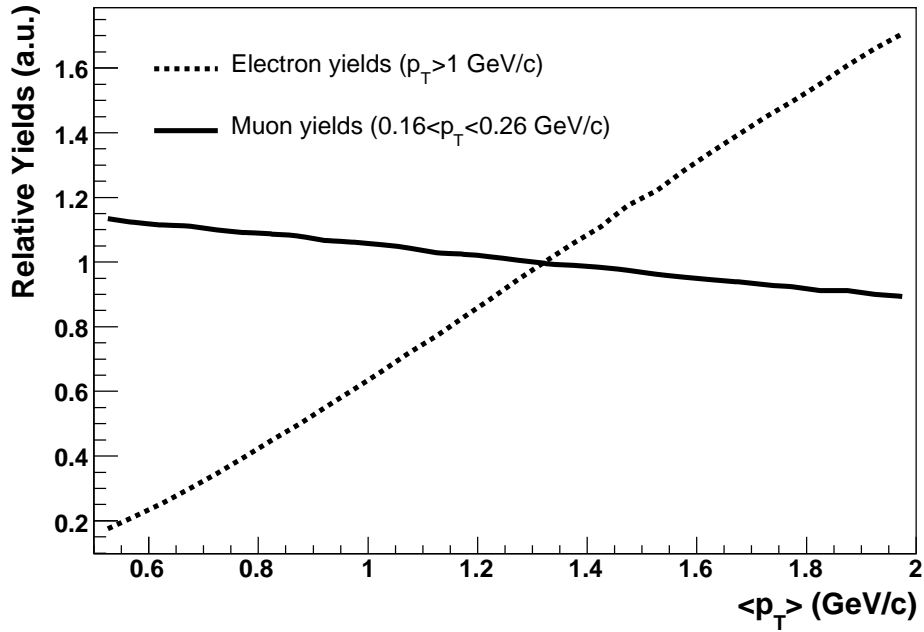}
  \caption{Lepton yields relative to the fixed total charm cross
   section as function of power-law parameters $\langle p_{T} \rangle$
   for a charmed hadron transverse momentum spectrum.  Solid line
   shows muon yields with a kinematics selection $0.16<p_T<0.26$
   GeV$c$ and $|y_l|<0.5$. Dashed line shows electron yields with
   $p_T>1.0$ GeV/$c$. }
  \label{fig:muonnmpt}
\end{figure}
The key issue then is how to identify and measure the charm lepton
spectrum at very low $p_T$ in the absence of high precision
measurements of directly reconstructed charmed hadrons. First, we will
demonstrate the feasibility by simple estimates and then using more
detailed Monte Carlo simulations from a combination of two types of
detectors: the Time Projection Chamber (TPC) and Time-of-Flight (TOF)
detectors.  

The ionization energy loss difference between $\pi$ and $\mu$ is about
12\% at $p=0.2$ GeV/$c$ and is much smaller at higher
momentum~\cite{pdgcharmff}.  Typical TPC ${\it dE/dx}$ resolutions are
$4\%$ at NA49/SPS, $8\%$ at STAR/RHIC and a projected $6\%$ at
ALICE/LHC. This means that the separation between $\pi$ and $\mu$ is
about 1.5$\sigma$ to $3\sigma$. Typical TOF timing resolution is about
$\sigma_t=100$ps at a distance of $l=2m$. Therefore, the separation of
$\pi$ and $\mu$ measured in time is: $\Delta t
=(\sqrt{m^{2}_{\pi}/p^{2}+1}-\sqrt{m^{2}_{\mu}/p^{2}+1})\times l\times
c\simeq600$ ps. This is about a 6$\sigma$ separation, which decreases
rapidly as $\gamma^{3}$ at higher momentum.

\begin{figure}
  \includegraphics[width=2.7in]{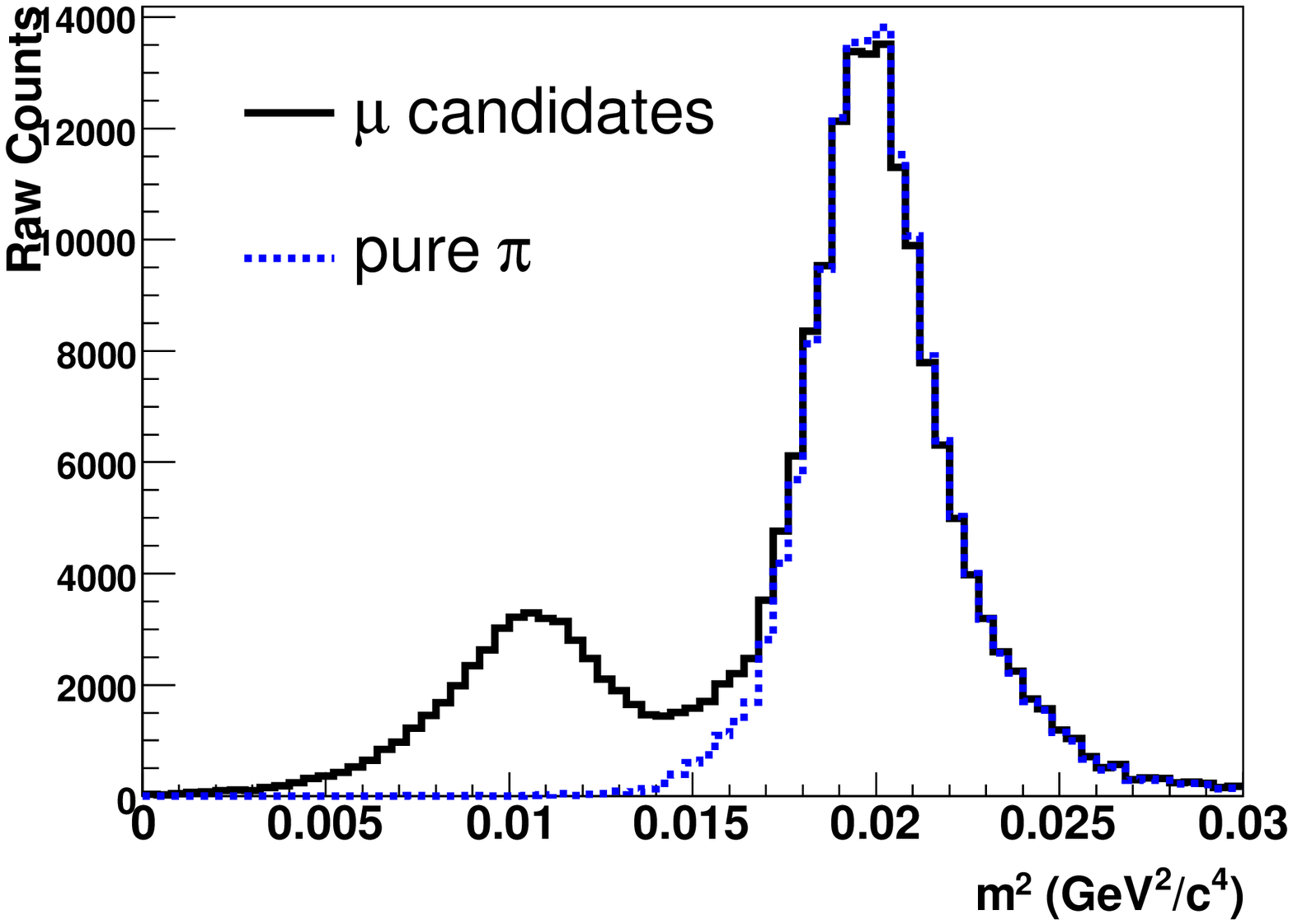}
  \includegraphics[width=2.7in]{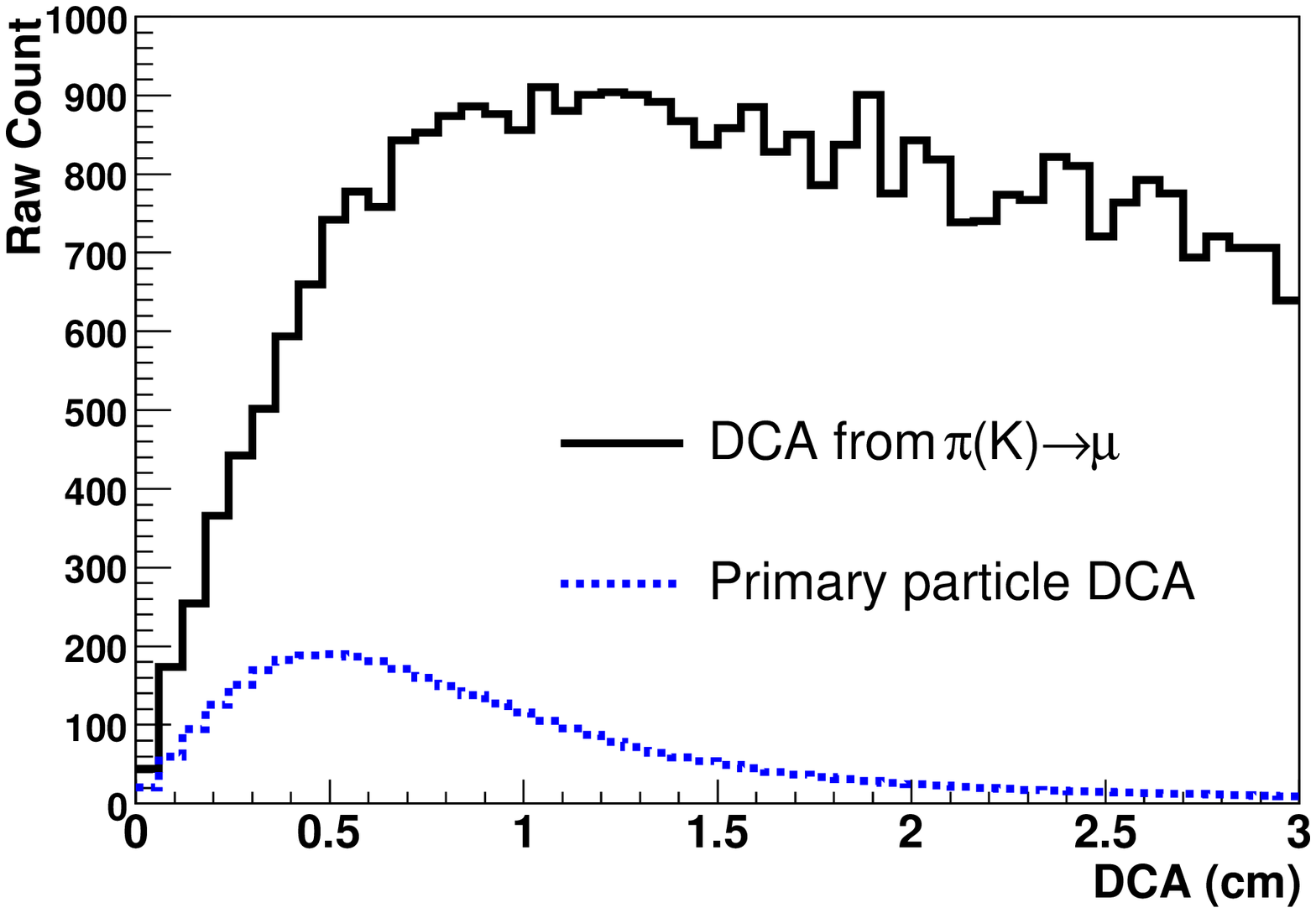}
  \caption{Left panel: Particle mass $m^{2}=(p/\beta/\gamma)^{2}$ from
Time-of-Flight detector after TPC {\it dE/dx} selections of muon
candidates (solid) and primary pion candidates (dashed).  Right panel:
Primary particle DCA distributon (dashed line) and muon DCA
distribution from background after TPC {\it dE/dx} and TOF $m^{2}$
selections from HIJING simulations through realistic STAR detector
configuration. The DCA distritions of muons from charmed hadron decay
are practically identical to the primary particle in this detector
configuration.}
  \label{fig:muDCA}
\end{figure}

For our more detailed simulations, we used the HIJING event
generator~\cite{hijing} to simulate Au+Au collisions in the STAR
detector at RHIC to demonstrate how muons can be identified and how
the background can be subtracted. The same approach can be used for
detectors with similar configurations, such as CDF and ALICE.  Muons
were identified by measuring the energy loss in the Time Projection
Chamber and the velocity in the Time-of-Flight patch at
STAR~\cite{pidNIMA,czhongSQM06}. From the {\it dE/dx} and
Time-of-Flight resolutions, we conclude that the {\it dE/dx}
difference between pions and muons is about $1.5\sigma$ and the mass
resolution is about $5\sigma$ at $p_T\simeq0.2$ GeV/$c$. This provides
a pion rejection factor of $>500$, $\sim25$ from each detector.  The
left panel of Fig.~\ref{fig:muDCA} shows the
$m^{2}=(p/\beta/\gamma)^{2}$ distribution from TOF after TPC {\it
dE/dx} selections~\cite{pidNIMA} for muon candidates and pure pion
candidates. A clean muon peak can be identified within a mass window
of $0.008<m^{2}<0.014$. The tail of the pion background can be
evaluated by selecting pure pion candidates with a {\it dE/dx} cut and
the result is shown as a dashed histogram in Fig.~\ref{fig:muDCA}. The
residual pions can be subtracted statistically from the distribution
of the distance of the closest approach (DCA) to the collision vertex
with the muon mass window applying to the pure pion
sample~\cite{czhongSQM06}.

The STAR Collaboration have performed analyses of all the V0 and other
weak-decay strange hadrons and has published many papers on these
topics~\cite{huilongPHD}. Those decay topologies have shorter decay
distances and comparisons between data and simulation show good
agreement.  Our analyses of DCA distributions are the similar and
should lead to a good match between data and
simulation~\cite{czhongSQM06}.  Measurements of the muon spectra from
charmed hadron semileptonic decays at low $p_{T}$ are not affected by
the $\pi^{0}$ Dalitz decays and photon conversions that an equivalent
electron measurement is.  The dominant background muons from pion/kaon
weak decays are subtracted using the DCA distribution. Other sources
of background
($\rho\rightarrow\mu^+\mu^-$,$\eta\rightarrow\gamma\mu^+\mu^-$, etc.)
are found to be negligible from the HIJING
simulation. Fig.~\ref{fig:muDCA} shows the DCA distribution of muons
from pion and kaon decays in minimum-bias Au+Au collisions. Also shown
is the DCA distribution of particles at the same $p_T$ originating
from the collision vertex. The integrated yields of these two
distributions reflect the appropriate muon(from charm)-to-hadron ratio
with hadron yields estimated from HIJING and the charm-to-muon yield
from d+Au collisions scaled by the corresponding $N_{bin}$. It is
clear that the two distributions are very different and muons from
charm decaycan be reliably obtained from this method with high
precision when the DCA distribution from real data is fitted to the
combination of these signal and background distributions. Since charm
production is expected to scale with $N_{bin}$ while low-$p_T$ pions
scale with number of the participant nucleons,$N_{part}$, we expect a
better signal-to-background ratio in central Au+Au
collisions. Additional vertex detectors, such as the Silicon Strip
Detector (SSD)~\cite{SSDNIM} and/or the Heavy Flavor Tracker
(HFT)~\cite{kaiQM05HFT}, can greatly enhance this capability because
the DCA of the muon from pion/kaon decay is determined by the decay
distance and track geometry while for the muon from charm decay, the
DCA accuracy is currently due to tracking resolution and can be
reduced to ${}^{<}_{\sim}100\mu m$ with a good vertex detector. In
principle, the 20\% systematic uncertainties presented by the STAR
Collaboration~\cite{czhongSQM06} can probably be further reduced. It
will then provide a complementary charm measurement in the era of
directly reconstructing charmed hadrons from a displaced secondary
vertex.

\begin{figure}
  \includegraphics[width=5in]{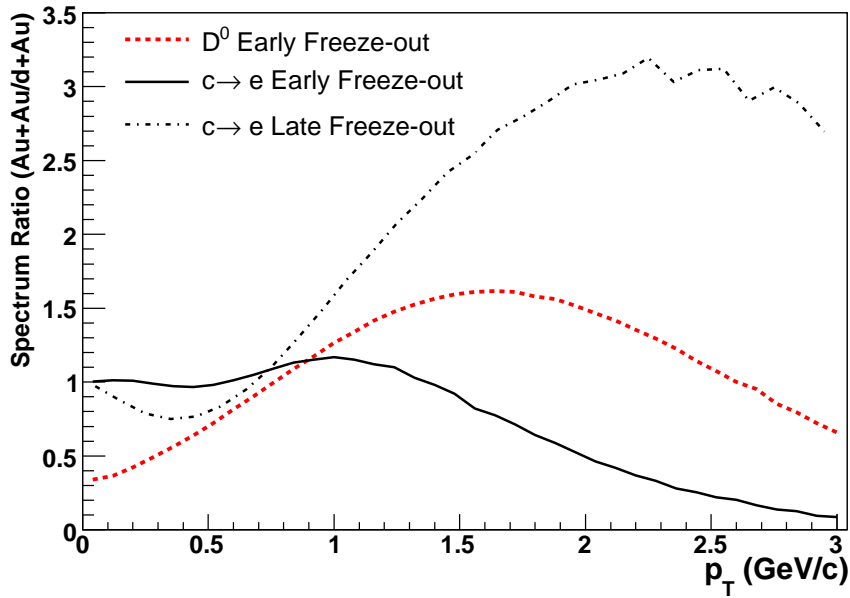}
  \caption{Charmed meson spectrum of blast-wave form with $T_{fo}=160$
MeV and $\beta_{max}=0.6$ (dashed line) divided by the spectrum with a
power-law parameterization of d+Au data from STAR~\cite{stardAucharm},
and corresponding electron spectrum ratio. These represent an early
freeze-out scenario. The dash-dotted line is for an electron spectrum
from a later freeze-out assumption with $T_{fo}=100$ MeV and
$\beta_{max}=0.9$.}
  \label{fig:BWpl}
\end{figure}

With the improved charm semileptonic decay form factor~\cite{cacciari}
and possible high precision measurements of muons at low $p_T$, we
re-evaluate the sensitivity of the lepton spectrum to the original
charm spectrum as has been performed previously~\cite{batsouli}. In
our approach, we take the extracted charm spectrum from d+Au
collisions as a baseline charm spectrum for nucleon-nucleon
collisions~\cite{stardAucharm}. A power-law function was fitted to the
$D^{0}$ $p_T$ spectrum combined with its associated decay electron
spectrum for the d+Au collision data~\cite{stardAucharm}. A blast-wave
similar to Ref.~\cite{batsouli} with freeze-out temperature and
freeze-out velocity parameters obtained from multistrange baryons
($T_{fo}=160$ MeV, $\beta_{max}=0.6$) was used to generate a charm
spectrum~\cite{multistrange}. The blast-wave assumption is likely not
appropriate for lepton $p_T>2$ GeV/$c$ and therefore we focused our
study for $p_T>2$ GeV/$c$. The spectrum was divided by the spectrum
from d+Au. Fig.~\ref{fig:BWpl} shows the ratios of these spectra to
the baseline spectra of charmed hadrons and leptons, respectively. The
semileptonic decay greatly smears the spectrum and reduces the
difference between the different spectrum shapes. However, it is clear
that a reasonably realistic blast-wave parameterization of charmed
mesons in Au+Au collisions is very different from that in d+Au
collisions. There is also a significant difference between spectra
with different flow (blast-wave function) parameters at $0.5<p_T<1.5$
GeV/$c$. Fig.6 shows that there is a factor of 3 difference at
$p_T=1.5$ GeV/$c$ between late freeze-out ($T_{fo}=100$ MeV,
$\beta_{max}=0.9$) and early freeze-out ($T_{fo}=160$ MeV,
$\beta_{max}=0.6$). Current measurements of non-photonic electron
spectra and direct charm spectra seem to be consistent with a
decreasing trend even at $p_T\simeq1.0$ GeV/$c$~\cite{starcharmQM05},
which is likely due to multiple collisions and thermalization at low
$p_T$ with early freeze-out~\cite{teaney} and not due to pQCD energy
loss. However, since the overall normalization is not known, an early
freeze-out scenario can be interpreted as suppression of the charm
total cross-section as well. This ambiguity can be resolved by a
measurement of the total cross-section.  In addition, the errors on
the current measurements are large in this $p_T$ range. We advocate
improving the measurements of electrons in this $p_T$ range to assess
if charm thermalizes in the medium and has similar flow and freeze-out
as multistrange hadrons~\cite{xu2}.

In summary, we propose a sensitive measurement of the charm total
cross-section at RHIC. The charm-decay muons at $p_T\simeq0.2$ GeV/$c$
can be identified by energy loss in a Time Projection Chamber and velocity
from a Time-of-Flight detector.  The background from pion/kaon decays can be 
subtracted using the distribution of the distance of the closest
approach to the collision vertex. We present simulations using 
HIJING within the realistic environment of the STAR detector. Detailed studies
show that charm-decay muon spectra at low $p_T$ are proportional to
the total charm cross-section. There is very weak dependence of the
muon yields at the level of $\pm15\%$ on charmed hadron shape over a
wide range of spectrum parameters. Detailed comparison of different
form factors of the charm semileptonic decay shows up to a factor of 1.5
difference in the resulting lepton spectra between PYTHIA and more
realistic form factors. A power-law spectrum and a blast-wave-type spectrum
possibly present in Au+Au is different by about $30\%$ at $p_T\simeq1$
GeV/$c$.  We conclude that the lepton spectrum at low $p_T$ from charmed
hadron decays is sensitive to the charm spectrum. Therefore, electron
spectra at $0.5<p_T<1.5$ GeV/$c$ can be used to study charm quark
thermalization and radial flow.

The authors thank STAR Collaboration for providing the detailed
detector configuration and simulations, Dr. Cacciari for providing a
parameterization of CLEO preliminary data, Drs. S. Blyth, X. Dong,
T. Ullrich, Nu Xu and H. Zhang for valuable discussions, and
Dr. S. Blyth for careful reading and comments of the manuscipt. This
work was supported in part by the HENP Divisions of the Office of
Science of the U.S. DOE; the Ministry of Education and the NNSFC of
China. XZB was supported in part by a DOE Early Career Award and the
Presidential Early Career Award for Scientists and Engineers.


\begin{thebibliography}{99}
\bibitem{cacciari} M. Cacciari, P. Nason and R. Vogt,
\Journal{\PRL}{95}{2005}{122001}
\bibitem{dokshitzer01} Y.L. Dokshitzer and D.E. Kharzeev,
        \Journal{\PLB}{519}{199}{2001}.

\bibitem{loic} L. Grandchamp and R. Rapp, \Journal{\PLB}{523}{60}{2001}.

\bibitem{pbm} A. Andronic {\it et al.},
        \Journal{\PLB}{571}{36}{2003}.


\bibitem{rafelski} R.L. Thews, M. Schroedter, and J. Rafelski,
        \Journal{\PRC}{63}{054905}{2001}.

\bibitem{mclerran} M.I. Gorenstein {\it et al.},
\Journal{\JPG}{28}{2151}{2002}.

\bibitem{matsui} T. Matsui and H. Satz,
        \Journal{\PLB}{178}{416}{1986}.

\bibitem{xu2} N. Xu and Z. Xu \Journal{\NPA}{715}{587c}{2003};
    Z.W. Lin and D. Molnar, \Journal{\PRC}{68}{044901}{2003};
    V. Greco, C.M. Ko and R. Rapp, \Journal{\PLB}{595}{202}{2004}.
\bibitem{teaney} G.D. Moore, D. Teaney, \Journal{\PRC}{71}{064904}{2005}

\bibitem{batsouli}
S. Batsouli {\it et al.}, \Journal{PLB}{557}{2003}{26-32}
e-print Arxiv: nucl-th/0212068

\bibitem{stardAucharm}
STAR Collaboration, J. Adams {\it et al.}, \Journal{\PRL}{94}{062301}{2005}
e-print Arxiv: nucl-ex/0407006

\bibitem{CDF}
CDF Collaboration, D. Acosta {\it et al.}, \Journal{\PRL}{91}{241804}{2003}

\bibitem{phenixAuAu}
PHENIX Collaboration, S.S. Adler, {\it et al.}, \Journal{\PRL}{96}{032301}{2006}
e-print Arxiv: nucl-ex/0510047
PHENIX Collaboration, S.S. Adler {\it et al.}, \Journal{\PRL}{94}{082301}{2005}
e-print Arxiv: nucl-ex/0409028
PHENIX Collaboration, K. Adcox {\it et al.}, \Journal{\PRL}{88}{192303}{2002} 
e-print Arxiv: nucl-ex/0202002

\bibitem{phenixpp}
PHENIX Collaboration, K. Adcox {\it et al.}, \Journal{\PRL}{96}{032001}{2006} 
e-print Arxiv: hep-ex/0508034 


\bibitem{miklos}
 S. Wicks, W. Horowitz, M. Djordjevic and M. Gyulassy,
e-print Arxiv: nucl-th/0512076

\bibitem{starcharmQM05}
Haibin Zhang {\it et al.}, Quark Matter 2005, Budapest, Hungary, 4-9 Aug. 2005;
e-print arXiv: nucl-ex/0510063; J. Bielcik {\it et al.}, e-print arXiv: nucl-ex/0511005

\bibitem{xindongQM05}
Xin Dong {\it et al.}, Quark Matter 2005, Budapest, Hungary, 4-9 Aug. 2005;
e-print Arxiv: nucl-ex/0509038

\bibitem{pdgcharmff} S. Eidelman {\it et al.}, \Journal{\PLB}{592}{1}{2004} (Particle Data Group); 
R.M. Barnett {\it et al.}, \Journal{\PRD}{54}{486}{1996}.

\bibitem{pythia} T. Sj\"ostrand {\it et al.}, \Journal{Computer Physics Commun.}{135}{238}{2001}.

\bibitem{cleocetalk} CLEO Collaboration, J. Yelton {\it et al.},
Presented at Heavy Quarks and Leptons, San Juan, Puerto Rico, June 4,
2004J.\\ http://www.lns.cornell.edu/public/TALK/2004/TALK04-42/ \\
N.E. Adam {\it et al.} (CLEO Collaboration), hep-ex/0604044 


\bibitem{polemass}G.S. Huang {\it et al.},
\Journal{\PRL}{94}{011802}{2005}; E.M. Aitala {\it et al.},
\Journal{\PRL}{80}{1393}{1998}.

\bibitem{xuISMD} Z. Xu, ISMD 2004, Acta Phys.Polon. {\bf B36} 243 (2005); 
e-print arXiv: nucl-ex/0410005

\bibitem{pidNIMA}
M. Shao {\it et al.}, \Journal{\NIMA}{}{}{2005} in press,
e-print Arxiv: nucl-ex/0505026; M. Anderson {\it et al.}, 
\Journal{\NIMA}{499}{659}{2003}

\bibitem{starwhitepaper}STAR Collaboration, J. Adams {\it et al.}, 
\Journal{\NPA}{757}{102}{2005}

\bibitem{hijing} X.N. Wang and M. Gyulassy, \Journal{\PRD}{44}{3501}{1991}.

\bibitem{czhongSQM06}
Haibin Zhang, Yifei Zhang, Chen Zhong for the STAR Collaboration, SQM06, LA, USA, Mar. 2006; 
Talks available at 
http://home.physics.ucla.edu/calendar/conferences/sqm2006/agenda/index.htm \\
A systematic uncertainty of 20\% has been presented in the final charm
total cross-sections that include the low momentum muons.

\bibitem{huilongPHD} Hui Long, Ph.D. Thesis, University of California
- Los Angeles, 2002;\\
http://www.star.bnl.gov/central/publications/theses/ \\
C. Adler {\it et al.},\Journal{\PRL}{89}{092301}{2002};
\Journal{\PRL}{89}{132301}{2002}; \Journal{\PLB}{595}{143}{2004};
J. Adams {\it et al.},\Journal{\PRL}{92}{052302}{2004};
\Journal{\PRL}{92}{182301}{2004}; \Journal{\PRL}{95}{122301}{2005}

\bibitem{SSDNIM}
L. Arnold {\it et al.}, \Journal{\NIMA}{499}{652}{2003}

\bibitem{kaiQM05HFT}
K. Schweda {\it et al.}, Quark Matter 2005, Budapest, Hungary, 4-9 Aug. 2005
e-print Arxiv: nucl-ex/0510003

\bibitem{multistrange} 
STAR Collaboration, J. Adams {\it et al.}, \Journal{\PRL}{92}{182301}{2004}

\end{thebibliography}
\end{document}